\documentclass[12pt]{article}
\input{epsf.sty}
\def\mydate{4 September 2001}
\def\ignore#1{{}}

\newcommand{\beeq}{\begin{equation}}
\newcommand{\eneq}{\end{equation}}
\newcommand{\beqn}{\begin{eqnarray}}
\newcommand{\eeqn}{\end{eqnarray}}

\def\mybig{\displaystyle \strut }

\def\la{\raise.16ex\hbox{$\langle$}\lower.16ex\hbox{}  }
\def\ra{\, \raise.16ex\hbox{$\rangle$}\lower.16ex\hbox{} }

\def\psibar{ \psi \kern-.65em\raise.6em\hbox{$-$} \lower.6em\hbox{} }
\def\AdS{{\rm AdS}}

\def\ep{\epsilon}

\def\myfrac#1#2{{\mybig #1\over \mybig #2}}

\begin{document}

%\rightline{\small OU-HET 356}
\rightline{\small \mydate}
\rightline{for J. Math. Phys.}

\vskip 4cm

\baselineskip=25pt

\centerline{\Large \bf Scaling Behavior in the Einstein-Yang-Mills}

\centerline{\Large \bf  Monopoles and Dyons}

\vskip 3cm

\centerline{\bf Yutaka Hosotani}

\centerline{{\small \it Department of Physics, Osaka University, Toyonaka,
Osaka 560-0043, Japan}\\
}

%\rightline{\small OU-HET 356}

\vskip 3.5cm

\begin{abstract}
Scaling behavior in the moduli space of  monopole and  dyon  solutions  in
the Einstein-Yang-Mills theory in the asymptotically anti-de Sitter 
space is derived.  The mass  of monopoles and dyons scales with respect to
their magnetic and electric charges, independent
of the values  of the cosmological constant and gauge coupling constant.
The stable monopole and dyon solutions are   approximated by 
solutions in the fixed anti-de Sitter spacetime.  
Unstable solutions  can be viewed as
the Bartnik-McKinnon solutions dressed with monopole and dyon solutions
in the fixed anti-de Sitter space.
\end{abstract}

%\pacs{PACS numbers: 04.40.Nr, 04.20.Jb, 11.27.+d}

\newpage

\baselineskip=20pt

%\tableofcontents
%\setcounter{page}{1}
\section{Introduction}

It has been shown that there exist a continuum of stable and
unstable monopole and dyon solutions in the Einstein-Yang-Mills theory in
the asymptotically  anti-de Sitter (AdS) space.\cite{BH1,BH2}  
They generalize a discrete family of unstable particle-like solutions in
the asymptotically Minkowski or de Sitter
space.\cite{Bartnik} Similarly, black hole solutions
exist with discrete values of magnetic charges in the asymptotically flat
or de Sitter space,\cite{Bizon,Volkov} and
with continuous values of  non-abelian electric and magnetic
charges in the asymptotically AdS space.\cite{Winstanley,BH2,Lugo,Torii}

Monopole and dyon solutions are characterized by their mass and
 non-Abelian magnetic/electric charges.  The spectrum defines 
the moduli space of the solutions, which varies with  the consmological
constant ($\Lambda$) and the gauge and gravitational coupling constants
($e$ and $G$). 
 The spectrum consists of infinitely many discrete points for $\Lambda \ge
0$, whereas it has a finite number of continuous branches for
$\Lambda <0$.   When the parameter $\Lambda < 0$ approaches zero,
an already-exsiting  branch of monopole and dyon solutions   collapses to a
single point in the  moduli space.  At the same time new branches of
solutions emerge.  A fractal structure in the moduli space
has been observed.\cite{BH2,Matinyan}

In this paper we derive a scaling law for the mass spectrum
 of the solutions with respect to their magnetic and
electric charges ($Q_M$ and $Q_E$), the cosmological constant $\Lambda(<0)$,
and the ratio of the gravitational constant to the gauge coupling constant
$v\equiv 4\pi G/e^2$.  Some of  the results in \cite{BH2} indicate that the
mass of  monopoles and dyons is expressed in terms of a universal function
$f(Q_M,Q_E)$. We shall show that this follows from the factorization
property of the  solutions and that $f(Q_M,Q_E)$ is determined by the
monopole and dyon solutions in the fixed AdS background metric.

AdS spacetime has many special properties.  In some models it accommodates
the holographic principle; the information on the  boundary of the space
determines physics in the bulk.\cite{Maldacena}  We shall see a trace of
this  property in the classical Einstein-Yang-Mills theory.  The existence
of stable monopole and dyon solutions in the asymptotically AdS space
seems tightly connected to boundary data on non-Abelian charges,
though more thorough investigation is necessary.

\section{Monopoles and dyons}

There exist static, spherically symmetric monopole and dyon solutions in
the Einstein-Yang-Mills theory.  The action of the system is 
\beeq
S=\int d^4x\sqrt{-g}  
\left[ \frac{1}{16\pi G}(R-2\Lambda)
- {1\over 4} F^{a\mu\nu} {F^a}_{\mu\nu} \right].
\label{action1}
\eneq
The Einstein and Yang-Mills equations are given by
\beqn
&R^{\mu\nu} - {1\over 2}g^{\mu\nu} (R - 2 \Lambda) = 8\pi G ~
T^{\mu\nu}&\cr
&{F^{\mu\nu}}_{;\mu} + e [A_\mu, F^{\mu\nu} ] = 0 ~~.&
\label{EYM-eq}
\eeqn
The metric of  spacetime is given by 
\beqn
&&\hskip -1cm
ds^2 = -\myfrac{H}{p^2} \, dt^2 + \myfrac{dr^2}{H} 
 + r^2 (d\theta^2 + \sin^2 \theta \, d\phi^2) \cr
&&\hskip -1cm
H = 1 - {2m\over r} - {\Lambda\over 3} \, r^2 
\label{metric1}
\eeqn
where $H$, $p$, and $m$ depend on $r$ only.
$m(r)/G$ represents a total mass contained inside $r$ in the $c=1$
unit.   (Numerical values below are given in the $c=\hbar=G=1$ unit.) 
The $SU(2)$ Yang-Mills fields take
\beeq
A^{(0)} = {\tau^j\over 2e} \Bigg\{ 
u(r)  {x_j\over r}  dt 
- \ep_{jkl} {1-w(r) \over r^2} { x_k} dx_l \Bigg\} 
\label{YM-ansatz1}
\eneq
in the Cartesian coordinates ($x_1^2+x_2^2+x_3^2=r^2$).  
The gauge coupling constant is denoted as $e$.
With these ansatz (\ref{metric1}) and (\ref{YM-ansatz1}) 
the Einstein and Yang-Mills equations (\ref{EYM-eq}) reduce to
\beqn
&&\left(\frac{H}{p}w^{\prime}\right)^{\prime} 
= -\frac{p}{H}u^2w-\frac{w}{p}\frac{(1-w^2)}{r^2} 
\label{YM1} \\
\noalign{\kern 5pt}
&&\left(r^2pu^{\prime}\right)^{\prime} = \frac{2p}{H}w^2u  
\label{YM2} \\
\noalign{\kern 5pt}
&&m' = v \bigg[
 H(w^{\prime})^2 + \frac{(1 - w^2)^2}{2r^2} 
+\frac{1}{2}r^2p^2(u^{\prime})^2 
+\frac{u^2w^2p^2}{H} \bigg] 
\label{Ein1}\\
\noalign{\kern 5pt}
&&p^{\prime} =
-\frac{2v}{r}p\left[(w^{\prime})^2+\frac{u^2w^2p^2}{H^2}\right] 
\label{Ein2} 
\eeqn
with the boundary conditions  $u=m=0$ and  $w=p=1$ at the origin.
The set of the equations contains two
parameters, the cosmological constant $\Lambda$ and 
the ratio of the gravitational constant to the gauge coupling constant 
$v={4\pi G}/{e^2}$.

There are soliton-type solutions with finite masses.  
There are infinitely many conserved, gauge-covariant charges.  In the 
spherically symmetric case the nonvanishing  charges of importance are
\cite{BH1,BH2}
\beqn
\pmatrix{Q_E\cr Q_M} &=& {e\over 4\pi} \int dS_k \,
 \sqrt{-g} ~ {\rm Tr}  \pmatrix{F^{k0}\cr \tilde F^{k0}}  
{x^j \tau^j\over r} \cr
&=&  \pmatrix{- u_1 p_\infty\cr 1 - w_\infty^2} ~~.
\label{charge1}
\eeqn
where $u_1$, $p_\infty$, and $w_\infty$ are defined by the asymptotic
expansion 
$u \sim u_\infty + (u_1/r) + \cdots$ etc.  
Each solution is specified by its mass (multiplied by $G$), $M=m(\infty)$, 
non-Abelian electric and magnetic charges, $Q_E$ and $Q_M$, and the number,
$n$,  of the nodes of $w(r)$.    For $\Lambda \ge 0$ the spectrum of the 
solutions is discrete, $u(r)=0$ ($Q_E=0$),   $n=1, 2, 3, \cdots$, and all
solutions are unstable.  For $\Lambda < 0$ the spectrum is completely 
different.  It is continuous.  For each $n$ ($=0, 1, 2, \cdots$) 
there are a family of solutions with continuous values of $Q_E$ and
$Q_M$.\cite{BH1,BH2}  In particular, the nodeless solutions ($n=0$) are
stable. In the moduli space of the solutions,   $M$ of a
particular point (solution) is a function of $\Lambda$, $v$, $n$, $Q_E$,
and 
$Q_M$.  $M$, $\Lambda$, and $v$ have dimensions of
(length), (length)$^{-2}$, and (length)$^2$, respectively, whereas
$n$, $Q_E$, and  $Q_M$ are dimensionless. We show that $M$ is expressed
in terms of a universal function of $Q_E$ and $Q_M$ up to an overall
factor.

\section{Solutions in the fixed AdS background metric}

To understand why stable solutions exist only in the asymptotically
AdS space ($\Lambda<0$), we consider soliton solutions
in the fixed AdS background metric, setting $p=1$ and $H=1-\Lambda r^2/3$
in Eqs.\ (\ref{YM1}) and (\ref{YM2}) and on the r.h.s.\ of 
Eq.\ (\ref{Ein1}).   Introduce $x=(|\Lambda|/3)^{1/2} r$ and
$\hat u = (3/|\Lambda|)^{1/2} u$.  Then Eqs.\ (\ref{YM1}),
(\ref{YM2}), and (\ref{Ein1}) become
\beqn
&&\hskip -1cm
{d\over dx} \bigg\{ (1+ x^2) {dw\over dx} \bigg\}
=  - \frac{w(1-w^2)}{x^2} -\frac{\hat u^2 w}{1+ x^2}\cr
\noalign{\kern 10pt}
&&\hskip -1cm
{d\over dx} \bigg\{ x^2{d\hat u\over dx} \bigg\}
  = \frac{2 w^2 \hat u}{1+ x^2}  
\label{fixed1} 
\eeqn
and
\beeq
{dm\over dx} = v \sqrt{ {|\Lambda|\over 3}} 
\Bigg\{  (1+x^2) \bigg( {dw\over dx} \bigg)^2
 + \frac{(1 -w^2)^2}{2x^2}
+\frac{x^2}{2} \bigg({d \hat u\over dx} \bigg)^2 
+\frac{\hat u^2 w^2}{1+x^2} \Bigg\} ~.
\label{fixed2} 
\eneq
The equations for $\hat u(x)$ and $w(x)$ do not involve either $v$ or
$\Lambda$. The charges of the solutions are
\beeq
Q_M = 1 - w_\infty^2 ~~,~~
Q_E = x^2 {d\hat u\over dx} \bigg|_{x=\infty} ~~.
\label{charge2}
\eneq
Hence a family of the solutions in the fixed AdS background metric
satisfy
\beqn
w(r; \Lambda,  w_\infty, Q_E)^\AdS &=& \tilde w^\AdS(x;w_\infty, Q_E) \cr
\noalign{\kern 5pt}
u(r; \Lambda,  w_\infty, Q_E)^\AdS &=& 
\sqrt{ {|\Lambda|\over 3}} \tilde u^\AdS(x;w_\infty, Q_E) ~~.
\label{fixed3}
\eeqn
Here $\{ \tilde w^\AdS(r), \tilde u^\AdS(r) \}$ represents a solution for
$\Lambda= -3$. Further $dm/dx$ is expressed in terms of $\hat u$ and $w$
with an overall  factor $v ( |\Lambda|/3)^{1/2}$, which implies that
\beqn
&&\hskip -1cm
m(r; \Lambda, v,  w_\infty, Q_E)^\AdS 
= v \sqrt{ {|\Lambda|\over 3}} \, \tilde m^\AdS(x; w_\infty, Q_E) \cr 
\noalign{\kern 5pt}
&&\hskip -1cm
M^\AdS = m^\AdS|_{r=\infty} 
= v \sqrt{|\Lambda|} f(Q_M, Q_E) ~~,
\label{ads1}
\eeqn 
where $\tilde m^\AdS(r)$ is the mass function for $v=1$ and $\Lambda=-3$.
$f(Q_M, Q_E) $ defines a  universal scaling function as we shall see below.
Note that $f$ is a double-valued function of $Q_M$ as $Q_M=1 - w_\infty^2$.

The size of the solutions also scales.  One definition of the size,
$\ell$,  of a solution is given in terms of $m(r)$ by
$m(\ell) = 0.5 \cdot m(\infty)$, 
where we have arbitrarily taken  a size-factor $0.5$.  It immediately
follows 
\beeq
\ell^\AdS = {1\over \sqrt{|\Lambda|}}  h (Q_M, Q_E) ~~.
\label{size1}
\eneq

\begin{figure}[tbh]\centering
\leavevmode 
\mbox{
\epsfxsize=7.5cm 
\epsfbox{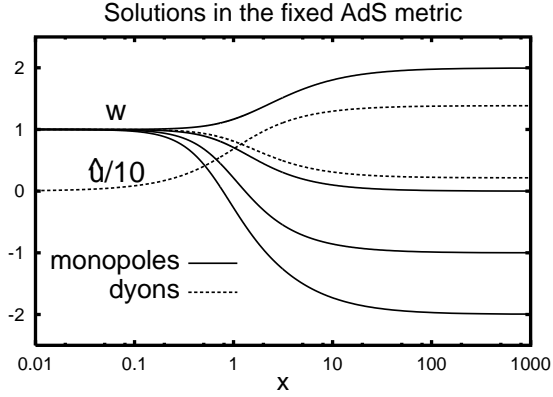}}  %\epsfbox{ads_w_u.eps}}
\caption{$w(x)$ and $\hat u(x)$ of typical monopole and dyon solutions in
the fixed AdS metric. The particular dyon solution displayed in the figure
has  $(Q_M, Q_E) = (0.954, 0.527)$.}
\label{ads-fig1}
\end{figure}

Typical solutions are depicted in fig.\ \ref{ads-fig1}.
Both $w(x)$ and $\hat u(x)$ monotonically decrease or increase.
The solutions have at most one node in $w(r)$.   The most of the energy
 of each solution is  localized in $x < 10$.

There is a special solution\cite{Brihaye1}
\beqn
&&\hskip -1cm
\hat u = 0 ~~,~~ w = {1\over \sqrt{1+x^2} } 
\label{ads-monopole1}
\eeqn
for which $Q_E = 0$, $Q_M = 1$, and 
$M = (\sqrt{3} \pi/ 8) \, v |\Lambda|^{1/2}$.
$Q_M=1$ corresponds to the same quantized magnetic
charge as for the 't Hooft and Polyakov monopole.

Further,  
$w^\AdS \sim 1$ and $m^\AdS \sim 0$ for $x < 0.1$.
It is also numerically confirmed that 
\beeq
{\rm Max}{\,}_r ~{2m^\AdS \over r} \cdot {1\over 1 - (\Lambda r^2/3)}
%&&\hskip .5cm 
\sim \cases{0.03 (v|\Lambda|)^{1/2} M^{1/2}|\Lambda|^{1/4} 
 &for $w_\infty > 1$\cr
0.1 M |\Lambda|^{1/2}  &for $w_\infty < 1$\cr}
\eneq
for monopole solutions.
As far as $v|\Lambda|$ and $M|\Lambda|^{1/2}$ are small enough,
corrections to the metric may be ignored, and the solution in the fixed
AdS background metric gives a good approximation to a solution in the
EYM theory.  ($p(r)\sim 1$ for those solutions.) 

\section{Factorization}

Let us turn to the EYM solutions in the $\Lambda=0$ case.
Set $u=0$.  Expressed in terms of $y=r/\sqrt{v}$, 
$H=1 - (2\bar m/y)$ and  $\bar m=m/\sqrt{v}$,
Eqs.\ (\ref{YM1}), (\ref{Ein1}), and (\ref{Ein2}) contain no parameter;
\beqn
&&\hskip -1cm
{d\over dy} \left(\frac{H}{p} \frac{dw}{dy} \right)
= -\frac{w}{p}\frac{(1-w^2)}{y^2}   \cr
\noalign{\kern 10pt}
&&\hskip -1cm
\frac{d \bar m}{dy} = 
\frac{(1 -w^2)^2}{2y^2}  +H \bigg( \frac{dw}{dy} \bigg)^2 \cr
\noalign{\kern 10pt}
 &&\hskip -1cm
\frac{dp}{dy} = -\frac{2p}{y} \bigg(  \frac{dw}{dy} \bigg)^2
\label{BKeq} 
\eeqn
  Solutions
$\{w, p, \bar m \}$ are functions of $y$ only.  In each
solution $w(r)$ crosses the axis $n$ times ($n=1,2,\cdots$), and
approaches $(-1)^n$ asymptotically.  A physical mass is given by
$M/G$, or $\bar m(\infty) \sqrt{v}/G
 = \bar m(\infty) M_{\rm Pl}/\sqrt{\alpha}$
where $\alpha=e^2/4\pi$ and $G= M_{\rm Pl}^{-2}$.  The mass of
the $n$-th Bartnik-McKinnon solution is  
\beeq
{({\rm mass})}_n^{\Lambda=0}(v) = {M_{\rm Pl}\over \sqrt{\alpha}} ~
e_n \hskip 1cm (n=1, 2, \cdots).
\label{mass1}
\eneq
$e_n = M_n^{\Lambda=0}|_{v=1}$ is numerically given by 
$(e_1, e_2,  \cdots)=( 0.8286 , 0.9713, \cdots)$.  For $n\gg 1$,
$e_n \sim 1 - 1.081 e^{-\pi n/\sqrt{3}}$.\cite{Breitenlohner1}
$w_n^{\Lambda=0}(r) $, $p_n^{\Lambda=0}(r)$, and $m_n^{\Lambda=0}(r)$ of
the $n$-th solution reach  their asymptotic values at $r \sim
a_n \sqrt{v}$ where 
$a_n  \sim 10^n$.\cite{Volkov}  The size of the Bartnik-McKinnon
solutions is characterized by $\ell_n^{\Lambda=0} \sim a_n \sqrt{v} $.

Monopole and dyon solutions in $\Lambda < 0$ are labeled by
$(n, v, \Lambda, w_\infty, Q_E)$.  
The index $n$ runs over 0, 1, 2, $\cdots$.   We would like to show that
for $ \ell_n^{\Lambda=0} \sqrt{|\Lambda|} \ll 1$, 
the Einstein-Yang-Mills monopole solutions are well approximated by
\beqn
&&
w_n  
= w_n^{\Lambda=0}(r;v) \, w^\AdS(r; \Lambda, (-1)^n w_\infty, Q_E) ~, \cr
&&
u_n 
= u^\AdS(r; \Lambda, (-1)^n w_\infty, Q_E) / p_n^{\Lambda=0}(\infty;v) ~,\cr
&&
p_n = p_n^{\Lambda=0}(r;v) ~,\cr
&&
m_n
= m_n^{\Lambda=0}(r;v) + m^\AdS(r; \Lambda, (-1)^n w_\infty,Q_E, v) ~,
\label{approx1}
\eeqn
where it has been understood that
$w_0^{\Lambda=0}(r;v)=p_0^{\Lambda=0}(r;v)=1$ and
$m_0^{\Lambda=0}(r;v)=0$.  First, the solution in the fixed AdS
metric approximately solves the EYM equations for $n=0$, as remarked
above. 

% Fig 2
\begin{figure}[tbh]\centering
\leavevmode 
\mbox{
\epsfxsize=8.0cm \epsfbox{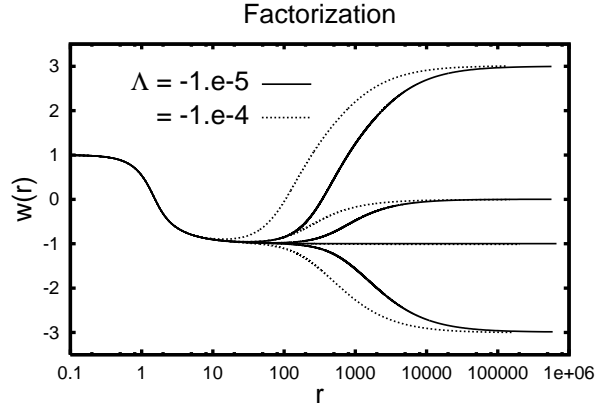}}
\caption{Factorization property of the EYM monopole solutions.
$\Lambda=-10^{-4}$ and $-10^{-5}$ with $v=1$.}
\label{factorize}
\end{figure}

Secondly, for $n \ge 1$ we consider two regions;   [I]  $r < 0.1(3/
|\Lambda|)^{1/2}$ and   [II]  $r> a_n \sqrt{v}$.  The two regions
overlap with each other if 
$|\Lambda| v < 0.03 \, a_n^{-2}$.  
In the  region I, 
$w^\AdS \sim 1$, $m^\AdS \sim 0$, and $-\Lambda r^2/3 \ll 1$
so that the solution is well approximated by that in the $\Lambda=0$
case, provided $u^2$ is sufficiently small. 
In the  region II, $w_n^{\Lambda=0} \sim (-1)^n$,
$p_n^{\Lambda=0} \sim p_n^{\Lambda=0}|_{r=\infty}$, and
$m_n^{\Lambda=0} \sim \sqrt{v} e_n$.  In Eqs.\ (\ref{YM1}) - (\ref{Ein2})
the value of constant $p$ is irrelevant with $pu$ substituted by $u$.
$H$ is approximated by $H=1 - \Lambda r^2/3$ as 
$m_n^{\Lambda=0}/r < 1/a_n \ll 1$.  Hence the solutions are given by
those in the fixed AdS background metric with 
$w$ at ${r=\infty} $ given by  $(-1)^n w_\infty$.
The solutions in the asymptotically AdS space are obtained by dressing 
solutions in the fixed-AdS background metric to the 
Bartnik-McKinnon solutions in the asymptotically flat space, as expressed
in (\ref{approx1}).

The factorization property of the solutions, (\ref{approx1}), is 
confirmed by numerical evaluation of the solutions.  In fig.\
\ref{factorize}, $w(r)$ of the monopole solutions at $\Lambda = - 10^{-4},
-10^{-5}$ and $v=1$ with various $w_\infty = 3, 0, -1, -3$ are depicted.
For $r <10$ these solutions are well approximated by the 
first Bartnik-McKinnon solution at $\Lambda=0$.  For larger $r$
the solutions are essentially given by 
$- w^\AdS (x; \Lambda=-1, -w_\infty)$.

\section{Scaling}

A scaling law follows from  the factorization property.  From
(\ref{ads1}) and (\ref{approx1})
\beeq
{ M_n(v,\Lambda, w_\infty,Q_E) - \sqrt{v} e_n \over v\sqrt{|\Lambda|}}
= f (Q_M,Q_E)  
\label{scaling1}
\eneq
the r.h.s.\ of which is independent of $\Lambda$ and $v$, and also of $n$. 
The scaling law is valid for $|\Lambda| v < 0.03 a_n^{-2}$
and small $|Q_E|$.

In fig.\  \ref{scaling-mass2} numerical data of
monopole solutions for the lowest ($n=0$) and second ($n=1$) branches is
depicted.  It is seen that all data for
$v |\Lambda| < 0.01$ falls on the universal function $f(Q_M,0)$ for
$n=0$, and  for $v|\Lambda| < 0.0001$ for $n=1$.
It follows from (\ref{scaling1}) that the mass  is given by 
\beeq
({\rm mass})_n(v,\Lambda, w_\infty,Q_E)
= {e_n\over \sqrt{\alpha}} \, M_{\rm Pl} 
 + {\sqrt{|\Lambda|}\over \alpha} \, f(Q_M,Q_E) ~~.
\label{mass2}
\eneq
In the lowest branch the magnitude of the mass is determined by
$\sqrt{|\Lambda|}/ \alpha$, whereas in the higher branches it is given by
$M_{\rm Pl}/ \sqrt{\alpha}$.

% fig 3
{
\begin{figure}[h]
\centering \leavevmode 
\mbox{
\epsfxsize=8.cm \epsfbox{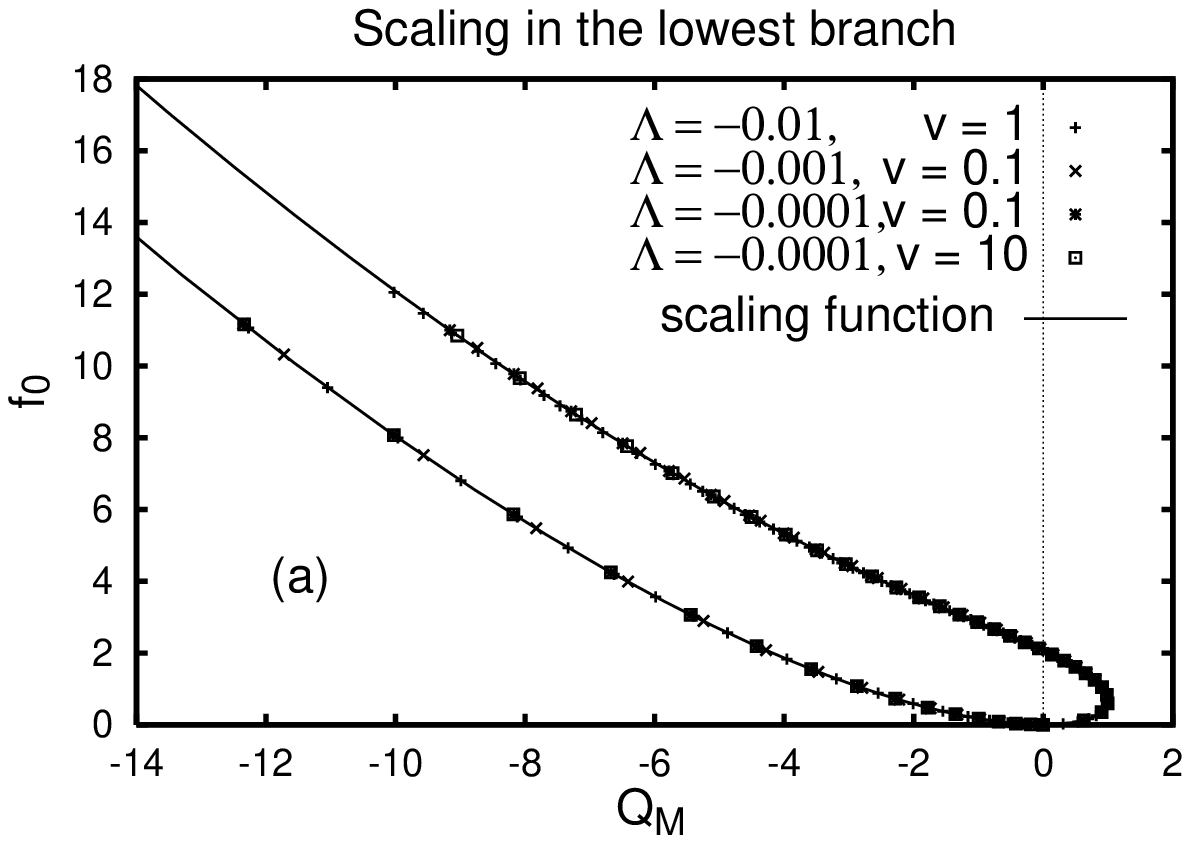}}
\end{figure}
%\vskip -.8cm
\begin{figure}[h]
\centering \leavevmode 
\mbox{
\epsfxsize=8.cm \epsfbox{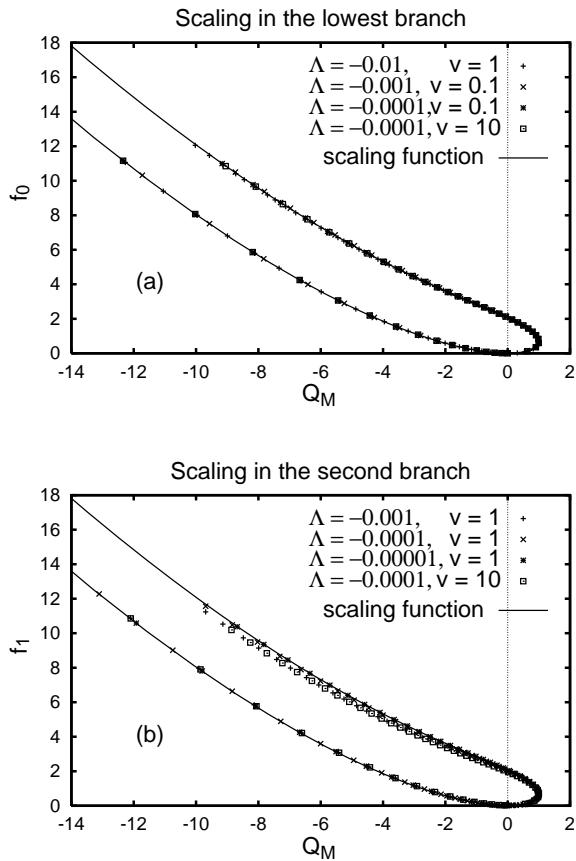}}
\caption{Scaling (a) for $f_0 \equiv  M_0(v,\Lambda,w_\infty,0) /(v
|\Lambda|^{1/2}) $,   and (b) for
$f_1 \equiv (M_1(v,\Lambda,w_\infty, 0)- v^{1/2} e_1)
  /(v |\Lambda|^{1/2})$.  The scaling function
$f(Q_M,0)$ is also depicted.}
\label{scaling-mass2}
\end{figure}
}

\begin{figure}[bht]
\centering \leavevmode 
\mbox{
\epsfxsize=8.cm \epsfbox{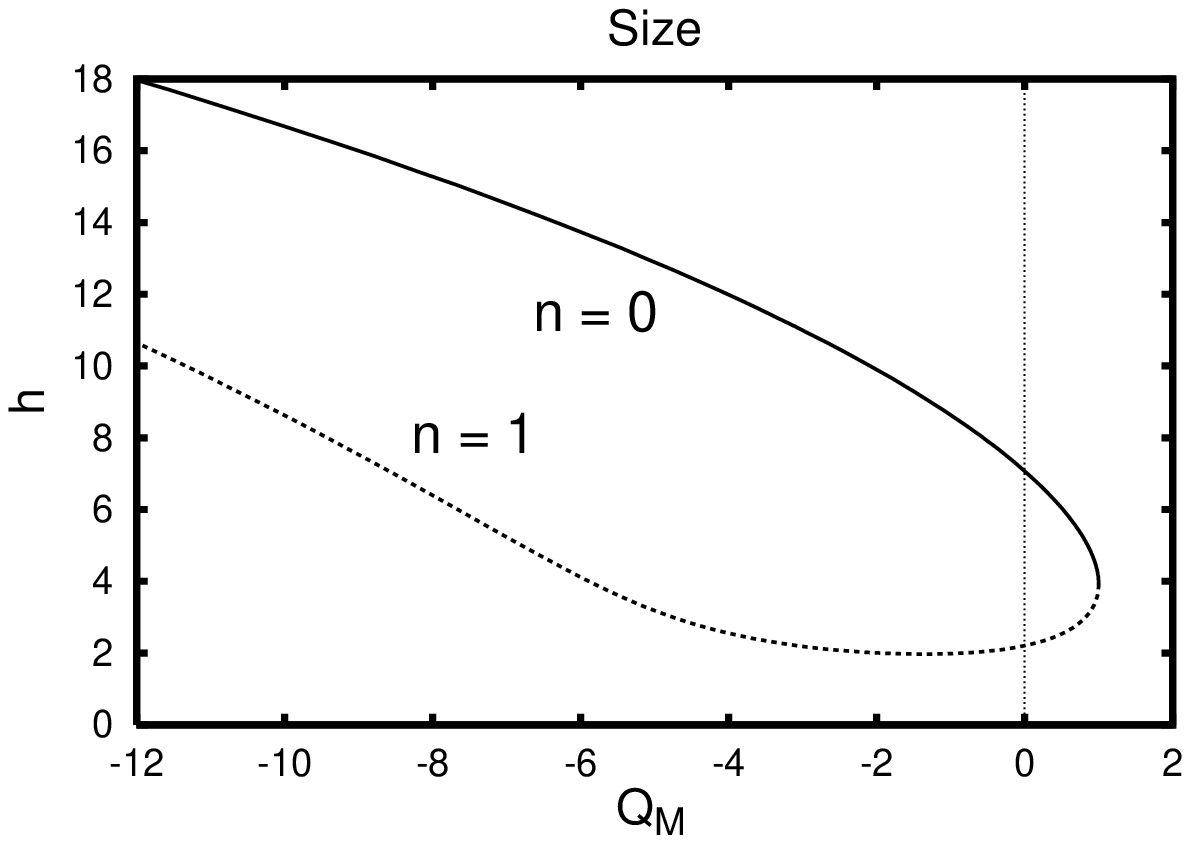}}
\caption{Scaling function, $h(Q_M, 0)$, for the size in (9). 
The portion
with $n=0$ or $n=1$ corresponds to solutions with no node or one node
in $w^\AdS(x)$, respectively.}
\label{scaling-size1}
\end{figure}

The size of a monopole or dyon is essentially the
same as that of the solution in the fixed AdS background metric, 
as the dressed fields cover  the inside Bartnik-McKinnon core.
$h(Q_M,0)$ in (\ref{size1}) is depicted in fig.\ \ref{scaling-size1}.

As $\Lambda (< 0)$ approaches 0, the branch in the
$Q_M$-$M$ plane collapses to a flat line  $M = \sqrt{v} e_n$.  
The size of the solutions grows as $|\Lambda|^{-1/2}$ so that 
Bartnik-McKinnon solutions with higher $n$ can be accommodated
inside the AdS  solutions, allowing  more solutions in higher branches. 
This explains the phenomenon observed in ref.\ \cite{BH2}.   In the 
$\Lambda=0$ limit only solutions with $Q_M=0$ survive.

\section{Summary}

In this paper we have examined monopole-dyon solutions in the 
Einstein-Yang-Mills theory in the asymptotically AdS space.
The monopole and dyon solutions  in the lowest branch ($n=0$) are
essentially the  solutions in the fixed AdS background metric.  The
solutions in  the higher branches ($n>0$) are obtained by dressing monopole 
and dyon solutions in the fixed AdS background metric around the
Bartnik-McKinnon solutions in the asymptotically  flat space.   As all
Bartnik-McKinnon solutions are unstable, the monopole and dyon solutions in
the higher branches are unstable, whereas the nodeless solutions are stable
against small perturbations.

Because of the factorization property of the solutions there
arises a scaling law in the mass of the solutions when regarded as
a function of $e$, $G$, $\Lambda$,  $Q_M$ and $Q_E$.  Up to an
overall factor it scales to a universal  function 
$f(Q_M,Q_E)$ determined by the solutions in the fixed
AdS metric.  The factorization/dressing mechanism is expected
to apply for black hole solutions as well.

In quantum theory $Q_M$ and $Q_E$ are expected to be quantized. 
Solutions with minimal $|Q_M|$ or $|Q_E|$ must be absolutely stable.
The stable solutions discussed in the present paper are, in nature,
non-topological solitons.  They exist only with gravitational force.
In this sense they may be called gravitational 
solitons.\cite{Weinberg}

% A useful Journal macro
%\def\jnl#1#2#3#4{{#1}{\bf #2} (#4) #3}
\def\jnl#1#2#3#4{{#1}{\bf #2}  #3 (#4)}
\def\Zphys{{\em Z.\ Phys.} }
\def\jssc{{\em J.\ Solid State Chem.\ }}
\def\jpsJ{{\em J.\ Phys.\ Soc.\ Japan }}
\def\ptps{{\em Prog.\ Theoret.\ Phys.\ Suppl.\ }}

\def\JMP{{\em J. Math.\ Phys.} }
\def\NPB{{\em Nucl.\ Phys.} B}
\def\NP{{\em Nucl.\ Phys.} }
\def\PLB{{\em Phys.\ Lett.} B}
\def\PL{{\em Phys.\ Lett.} }
\def\PRL{\em Phys.\ Rev.\ Lett. }
\def\PRB{{\em Phys.\ Rev.} B}
\def\PRD{{\em Phys.\ Rev.} D}
\def\PRe{{\em Phys.\ Rep.} }
\def\AP{{\em Ann.\ Phys.\ (N.Y.)} }
\def\RMP{{\em Rev.\ Mod.\ Phys.} }
\def\ZPC{{\em Z.\ Phys.} C}
\def\SCI{\em Science}
\def\CMP{\em Comm.\ Math.\ Phys. }
\def\MPLA{{\em Mod.\ Phys.\ Lett.} A}
\def\IJMPB{{\em Int.\ J.\ Mod.\ Phys.} B}
\def\PR{{\em Phys.\ Rev.} }
\def\cmp{{\em Com.\ Math.\ Phys.}}
\def\JPA{{\em J.\  Phys.} A}
\def\CQG{\em Class.\ Quant.\ Grav. }
\def\ATMP{{\em Adv.\ Theoret.\ Math.\ Phys.} }
\def\ibid{{\em ibid.} }

\end{document}